\documentclass[final]{raa}            

\usepackage{graphicx,times}             
\usepackage{natbib}
\usepackage{amssymb,amsmath}
\bibpunct{(}{)}{;}{a}{}{,}

\usepackage[pagebackref=true]{hyperref}

\begin{document}

  \title{Dreicer electric field definition and runaway electrons in solar flares
}

   \volnopage{Vol.0 (20xx) No.0, 000--000}      
   \setcounter{page}{1}          

   \author{Yu.T. Tsap 
      \inst{1}
   \and A.V. Stepanov
      \inst{2}
   \and Yu.G. Kopylova
      \inst{2}
   }

   \institute{Crimean Astrophysical Observatory of the Russian Academy of Sciences,
                   Nauchny, 298409, Russia; {\it yuratsap2001@gmail.com}\\
        \and
             Pulkovo Observatory of the Russian Academy of Sciences,
             Pulkovskoje Shosse, 65/1, St. Petersburg, 196140, Russia\\
\vs\no
   {\small Received 20xx month day; accepted 20xx month day}}

\abstract{ We analyse electron acceleration  by a large-scale
electric field $E$ in a collisional hydrogen plasma under the
solar flare coronal conditions  based on  approaches proposed by
Dreicer and Spitzer for the dynamic friction force of electrons.
The Dreicer electric field $E_{Dr}$ is determined as a critical
electric field at which the entire electron population runs away.
Two regimes of strong ($E \lesssim E_{Dr}$) and weak ($E \ll
E_{Dr}$) electric field  are discussed. It is shown that the
commonly used formal definition of the Dreicer field leads to an
overestimation of its value by about five times. The critical
velocity at which the electrons of the ``tail'' of the Maxwell
distribution become runaway under the action of the sub-Dreiser
electric fields turns out to be underestimated by $\sqrt{3}$ times
in some works because the Coulomb collisions between runaway and
thermal electrons are not taken into account. The electron
acceleration by sub-Dreicer electric fields generated in the solar
corona faces difficulties. }

   \authorrunning{Yu.T. Tsap, A.V. Stepanov, Yu.G. Kopylova }            
   \titlerunning{Dreicer electric field }  

   \maketitle

%
%
\section{Introduction}           
\label{sect:intro}

Solar flares  are a conversion process of  free magnetic energy to
kinetic and thermal energy. Moreover, they are a major particle
accelerator in the solar system \citep{Reames15}. Almost all
electrons contained in flare coronal loops should be accelerated
\citep{Miller etal.97}. This suggests that the very effective
electron acceleration mechanism should be implemented during the
flare energy release, for example, associated with the
 large-scale electric field generation \citep{Zaitsev
etal.16, Fleishman etal.22}.

In a fully ionized plasma, the collisional friction force is
inversely proportional to the square of the electron velocity $v$
 if  it exceeds the most probable thermal velocity $v_{Te}$
\citep[see, e.g.,][]{Trubnikov65}. As a result, a strong electric
field acceleration force can overcome the collisional damping,
accelerating high energy (runaway) electrons to relativistic
speeds. The Dreicer electric field is the fundamental concept of
this phenomenon \citep{Dreicer58,Dreicer59, Harrison60,
Trubnikov65, Gurevich78, KnoepfelSpong79, Kaastra83, Benz02,
Aschwanden04, Bellan06, Zhdanov et al.07, Fleishman13,
Marshall19}. According to the generally accepted definition, the
Dreicer electric field $E_{Dr}$ (or Dreicer field) is a critical
electric field at which electrons in a collisional plasma with $v
\approx v_{Te}$ can be accelerated, i.e., the entire electron
population runs away \cite[e.g.,][]{Holman85}. The field $E_{Dr}$
was named after Harry Dreicer who derived the corresponding
expression for the critical electric field in 1958
\citep{Dreicer58,Dreicer59}.

For the first time, the  idea of runaway electrons was outlined by
the Nobel Prize laureate Wilson \citep{Wilson24} to explain
thunderbolts in the Earth's atmosphere and  was further developed
by Gurevich \citep[e.g.,][]{GurevichZybin01}.  Gurevich's theory
was applied by~\citet{Tsap etal.22} \citep[see also][]{Tsap
etal.2020} in relation to the acceleration of electrons in the
lower solar atmosphere during flares. However, the origin of
strong electric fields was not considered.

The mechanism for ion runaway is different from electron runaway
\citep[e.g.,][]{Gibson59, Gurevich61, Furth72, Holman95,
Fleishman13}.
The positive test charge experiences two opposite forces:
acceleration due to $E$, and friction with the moving electrons.
If the test charge has the same charge as the bulk ions, these two
forces must be equal and opposite when the electric field $E <
E_{Dr}$. However, if the ionic charge $Z$ differs from the charge
of bulk ions $Z_b$, the forces scale differently with $Z$:
electric field acceleration scales as $Z$, while friction on the
drifting electrons scales as $Z^2$. Therefore, for $Z >Z_b$, the
dominant force on the test charge will be electron friction, and
the charge will be dragged to high energies as its velocity
equilibrates with the electron mean flow. For $Z < Z_b$, friction
becomes unimportant, so the test charge accelerates along $E$.
Note that the total drag force on an ion does not monotonically
fall off below  $v_{Te}$, but has a minimum and in a multispecies
plasma ''partial runaway'' can occur. As to the solar flares,
\citet{Holman95} has shown that the ions will be freely
accelerated to energies greater than $\sim 1$~MeV only if they are
able to overcome the electron drag or if the entire electron
population is freely accelerated, i.e., the electric field exceed
the Dreicer field.

Despite the concept of the Dreicer electric field is quite common
in solar physics, there are some essential inconsistences. In
particular, the formulae for the Dreicer electric field can differ
by a factor of 4.7 (e.g., Aschwanden 2004, Equation 11.3.2; Bellan
2006, Equation 13.85).
Therefore, this issue
requires a more detailed analysis.

The purpose of this work is to clarify the reason for the existing
inconsistences and to discuss the consequences of the results in
the light of the electron accelerations in solar flares.

\section{Dynamic Friction Force and Dreicer Electric Field}
\label{sect:dynfric}

Let us consider two regimes in the motion of electrons under the
action of the electric field. In the limit of the strong field
regime ($E \lesssim E_{Dr}$), the encounters between alike
particles do not contribute to dynamical friction. In the weak
field regime ($E \ll E_{Dr}$), as distinguished from the previous
case, the acceleration of runaway electrons is possible only in
the ``tail'' of the Maxwellian distribution function, and we take
into account the Coulomb collisions of accelerated electrons not
only with thermal  ions but also with thermal electrons of the
background plasma.

\subsection{Strong field regime}

Following Dreicer \citep{Dreicer58,Dreicer59} \cite[see
also][]{Trubnikov65} for the Maxwellian distribution function of
electrons at the initial moment of time and ion gas with zero
temperature,  neglecting the interaction between alike particles
and using the standard notation, the solution of the Boltzmann
equation
\begin{linenomath}
\begin{equation}
\frac{d f}{dt} + \frac{e{\bf E}}{m} \frac{d f}{d {\bf V}} =
\left.\frac{d f}{d t}\right|_c,
\end{equation}
\end{linenomath}
can be represented as the displaced Maxwellian distribution
function
\begin{linenomath}
\begin{equation}
f(t, v) = n
\left(\frac{m}{2kT}\right)^{3/2}\exp\left\{-\frac{m}{2kT} ({\bf V}
- {\bf v}(t))^2 \right\}.
\end{equation}
\end{linenomath}
Here the average electron velocity ${\bf v}(t)$ is the solution of
an equation of motion which includes the effects of collisions and
has the form
\begin{linenomath}
\begin{equation}
\label{motion}
 m\frac{dv}{dt}  = e(E - E_c  G(x)), \;\;\; x = \frac{v}{v_{Te}},
 \;\;\; v_{Te}= \sqrt{2kT/m},
\end{equation}
\end{linenomath}
where the Chandrasekhar function
\begin{linenomath}
$$
G(x) = \frac{\Phi(x) - x \Phi'(x)}{2x^2},\;\;\; \Phi(x) =
\frac{2}{\sqrt{\pi}} \int_0^{x} e^{-y^2}dy,
$$
\end{linenomath}
and the critical electric field $E_c$  is
\begin{linenomath}
\begin{equation}
\label{Dreicer Ec} E_c = \frac{4 \pi n e^3}{kT} \ln \Lambda.
\end{equation}
\end{linenomath}
Note that Dreicer \citep{Dreicer58,Dreicer59} used the function
$\Psi(x) = 2 G(x)$ instead of  $G(x)$.

It should be stressed that the electric field $E_c$ (Norman \&
Smith 1978; Holman 1985; de Jager 1986; Benz 2002, Equation 9.2.6;
Aschwanden 2004, Equation 11.3.2; Tsap \& Kopylova 2017)
or $E_c/2$ \citep{Kuijpers et al.81,Moghaddam-Taaheri90} commonly
called the Dreicer electric field  $E_D$ in the papers devoted to
the electron acceleration in solar flares. Meanwhile, the
Chandrasekhar function $G(x)$ reaches the maximum at  $x \approx
1$ ($v \approx v_{Te}$) and $G(1) \approx 0.214$ \citep[see,
e.g.,][]{Trubnikov65}. As a result, the
 condition of the acceleration of runaway electrons with $v \approx v_{Te}$, in view of
 Equations~(\ref{motion}) and (\ref{Dreicer Ec}),
takes the form (see also, Dreicer 1958; Dreicer 1959; Trubnikov
1965;  Golant et al. 1977, Equation 7.174; Bellan 2006, Equation
13.85)
\begin{linenomath}
\begin{equation}
\label{Dreicer ED}
 E  > E_{min} = E_{Dr} =
E_c G(1)\approx \alpha \frac{4 \pi n e^3}{kT}  \ln \Lambda =
\alpha \frac{e}{r_{De}^2} \ln \Lambda,
\end{equation}
\end{linenomath}
where $\alpha \approx 0.214$ and the Debye radius $r_{De} =
\sqrt{kT/(4 \pi ne^2)}$. The inequality, $E > E_{Dr}$, can be
considered as a condition for runaway acceleration when all
electrons accelerate to  high energy.

It is interesting to note that sometimes for the definition of the
Dreicer electric field kinetic effects connected with the velocity
distribution functions of charged particles are not taken into
account and the thermal electron velocity $\bar{v}_{Te} =
v_{Te}/\sqrt{2}$ instead of the most probable one $v_{Te}$ is used
\citep[e.g.][]{Holman85, Tsap2017}. In this case, $x=1/\sqrt{2}
\approx 0.71$ and the Chandrasekhar function $G(0.71)\approx
0.198$ \citep{Spitzer62}. Since $G(1) > G(0.71)$,
 the acceleration of the entire
electron population is formally impossible in this case because,
according to Equation~(\ref{motion}), the braking force
$$
F_D(x) = e E_c G(x),
$$
reaches the maximum at $x \approx 1$.

Thus, if we proceed from the definition that the Dreicer electric
field $E_{Dr}$ is the minimum electric field $E_{min}$ above which
electrons undergo free acceleration, then the Dreicer field
$E_{Dr} = E_{min}$. This approach seems to be more justified than
the approach based on $E_D = E_c$ and agrees with the definition
of the Dreicer electric field  $E_{Dr}$ proposed in (Bellan 2006,
Equation 13.85; Zhdamov et al. 2007, Equation 1.107; Marshall \&
Bellan 2019).
The commonly used formal Dreicer electric field is (Holman 1985;
Benz 2002, Equation 9.2.7; Aschwanden 2004, Equation 11.3.2)
\begin{linenomath}
\begin{equation}
\label{Dreicer mistake} E_D =  \frac{4 \pi n e^3}{kT} \ln \Lambda
= \frac{e}{r_{De}^2} \ln \Lambda,
\end{equation}
\end{linenomath}
and it turns out to be approximately $4.7$ times greater than the
Dreicer electric field~$E_{Dr}$ because, according to
Equations~(\ref{Dreicer ED}) and (\ref{Dreicer mistake}), the
ratio $E_{Dr}/E_D =\alpha$.

The obtained difference  is partially caused  by  different
approaches which are used for the dynamic friction force
calculation. For example,  according to Spitzer~\citep{Spitzer62},
the dynamic friction force for the electron flux (test particle)
caused by the Coulomb collisions with Maxwellian thermal protons
is (Harrison 1960; Spitzer 1962, Equation 5.15; Knoepfel \& Spong
1979)
\begin{linenomath}
\begin{equation}
\label{force F_p protons}
 F_{ep} = \frac{4 \pi n e^4}{kT} \ln
\Lambda \frac{M}{m} G(\sqrt{M/m}x),
\end{equation}
\end{linenomath}
where  $M$ is the mass of a proton.

Assuming $\sqrt{M/m}x \gg 1$ ($G(y \gg 1) \approx 0.5y^{-2}$),
instead of Equation~(\ref{force F_p protons}), we have
\begin{linenomath}
\begin{equation}
\label{fep}
 F_{ep} = \frac{4 \pi n e^4}{mv^2} \ln \Lambda.
\end{equation}
\end{linenomath}
Note that the square root of $\sqrt{M/m}\approx 43$ and  for the
electron velocity $v=v_{Te}$ ($x = 1$) from
Equations~(\ref{Dreicer mistake})-(\ref{fep}) we find the
``Dreicer electric field''
\begin{linenomath}
\begin{equation}
\label{EDS}
 E_{DS} = \frac{F_{ep}}{e} = \frac{2 \pi n e^3}{kT} \ln
\Lambda = \frac{E_D}{2}.
\end{equation}
\end{linenomath}
Equation (\ref{EDS})  agrees with the appropriate expressions in
\citep{Kuijpers et al.81,Moghaddam-Taaheri90}.

 It should be stressed that ~\citet{Spitzer62} did not take into
consideration the velocity distribution of
 electrons  exposed to an external electric field. In spite of that Spitzer's  and Dreicer's
approaches for dynamic friction forces give the same results at $x
= v/v_{Te}  \gg 1$ because from Equations~(\ref{motion}),
(\ref{Dreicer Ec}), and  (\ref{fep}) we have
\begin{linenomath}
$$
F_D = eE_c G(x \gg 1)  = \frac{4 \pi n e^4}{mv^2} \ln \Lambda =
F_{ep}.
$$
\end{linenomath}

In fact, in accordance with Equation~(\ref{force F_p protons}),
the friction force $F_{ep}$ reaches the maximum value when the
electron velocity $v$ is equal to the thermal proton velocity
$v_{Tp} = \sqrt{2kT/M}$ ($\sqrt{M/m}x =1$) and
\begin{linenomath}
$$
F_{ep}^{max} = \alpha\frac{4 \pi n e^4}{kT} \frac{M}{m} \ln
\Lambda.
$$
\end{linenomath}
Since  $F_{ep}^{max} \gg F_D(x=1)$,  we can conclude that
Spitzer's approach does not work for slow ($v \ll v_{Te}$)
electrons in the strong field regime \citep[see also
Fig.~1,][]{Holman95}.

\subsection{Weak field regime}

In the general case,  \citet{Spitzer62} has shown that the total
dynamic friction force for the electron flux with the same initial
velocity  due to the Coulomb collisions with thermal electrons and
protons of a Maxwellian hydrogen fully ionized plasma  is
(Harrison 1960; Spitzer 1962, Equation 5.15; Knoepfel \& Spong
1979)

\begin{linenomath}
\begin{equation}
\label{forceFp}
 F_S = \frac{4 \pi n e^4}{kT} \ln
\Lambda\{2G(x) + \frac{M}{m} G(\sqrt{M/m}x)\},
\end{equation}
\end{linenomath}
where  the first  term on the right-hand side of
Equation~(\ref{forceFp}) corresponds to the friction force  caused
by electron-electron collisions $F_{ee}$.  Then it follows from
Equation~(\ref{forceFp}) that at  $x \gg 1$ we have (Golant et al.
1977, Section 7.11)
\begin{linenomath}
\begin{equation}
\label{force Fp approx}
 F_S \approx F_{ee} + F_{ep} = \frac{12 \pi n e^4}{mv^2} \ln
\Lambda,
\end{equation}
\end{linenomath}
where
\begin{linenomath}
$$
F_{ee} = \frac{8 \pi n e^4}{mv^2} \ln \Lambda\,.
$$
\end{linenomath}
This allows us to find the critical velocity $v_{cr}$ for runaway
electrons based on the equality between the electric and the
dynamic friction  force, which has the form
\begin{linenomath}
$$
F_S = eE.
$$
\end{linenomath}
Consequently, using Equation~(\ref{force Fp approx}), we get
\begin{linenomath}
\begin{equation}
\label{critical velocity}
 v_{cr}^2 = \frac{12 \pi n e^3}{mE} \ln \Lambda\,.
\end{equation}
\end{linenomath}
Equation~(\ref{critical velocity}) agrees well with the
appropriate expression in \citep[][Equation 7.176]{Golant77}.
After that, in view of Equations~(\ref{Dreicer mistake}) and
(\ref{critical velocity}), we find
\begin{linenomath}
\begin{equation}
\label{critical velocity+Ed} v_{cr}^2 = \frac{3}{2}
\frac{E_{D}}{E} v_{Te}^2.
\end{equation}
\end{linenomath}

It should be stressed that according to \citet{KnoepfelSpong79}, the square of the critical velocity
is
\begin{linenomath}
\begin{equation}
\label{critical velocity old}
 v_{c}^2 =  \frac{4 \pi n e^3}{mE} \ln \Lambda  = \frac{E_D}{2E} v_{Te}^2.
 \end{equation}
 \end{linenomath}
 Comparing Equations~ (\ref{critical velocity+Ed}) and
 (\ref{critical velocity old}), it easy to conclude that the critical velocity in \citep{KnoepfelSpong79} was underestimated
by $\sqrt{3}$ times
 ($v_{cr}/v_c = \sqrt{3}$) because authors did not take into account collisions
between runaway and thermal electrons as distinguished from us and \citet{Golant77}.

A small difference between values of  $v_{cr}$ and $v_c$ can be
very important to estimate the number of runaway electrons in the
``tail'' of the Maxwellian  distribution function. Indeed, as it
follows from (Kaplan \& Tsytovich 1972, Equaton 9.10; Holman 1985)
the ratio of the
accelerated electrons to their total number is
\begin{linenomath}
\begin{equation}
\label{kaplan}
 \frac{n_r}{n_e} \approx
\exp\left[-\left(\frac{v_r}{v_{Te}} \right)^2\right].
\end{equation}
\end{linenomath}
Then, from Equation (\ref{kaplan}) we derive
\begin{linenomath}
\begin{equation}
\label{estimate}
 \frac{n_{cr}}{n_c} \approx
\exp\left[-\frac{v_{cr}^2-v_c^2}{v_{Te}^2}\right] =
\exp\left[-\frac{2v_{cr}^2}{3 v_{Te}^2}\right].
\end{equation}
\end{linenomath}
Supposing $v_{cr} = 3 v_{Te}$, we find from
Equation~(\ref{estimate}) that $n_{cr}/n_c = 2.5\times 10^{-3}$
because the total friction force $F_S$ is greater than $F_{ep}$.
Therefore, the difference in the number of runaway electrons can
reach orders of magnitude in spite of the small difference between
values of $v_{cr}$ and~$v_c$. This means that the electron
acceleration in solar flare coronal loops by sub-Dreicer electric
fields faces difficulties \cite[see for details][]{Tsap etal.22}.

\section{Discussion and Conclusion}
\label{sect:discussion}

We have shown that the definitions of the Dreicer electric field
differ in diverse works. This is partly explained by different
approaches proposed by \citet{Dreicer58, Dreicer59} and
\citet{Spitzer62}. In particular, Dreicer considered the
interaction between the electrons with the displaced Maxwellian
distribution and an ion gas at zero temperature, while Spitzer
investigated the evolution of the electron flux with the same
initial velocity in the Maxwellian plasma. These approaches
complement each other, but Equation~(\ref{Dreicer ED}) for the
Dreicer electric field $E_{Dr}$ seems to be the most adequate
because the distortion of the distribution function of electrons
under the action of electric field is taken into account in this
case. Note that some authors are restricted to the approximation
of pair collisions and do not take into account the kinetic
effects connected with the velocity distribution of charged
particles \citep[e.g.,][]{Tsap2017}.

The energy of runaway electrons can be essentially different
because of different definitions of the Dreicer electric field.
This may be  quite important point for electron acceleration in
solar flares. Indeed, the Dreicer electric field can be considered
as a rough estimate of the peak electric field  in the  coronal
 collisional  plasma. This suggests that the maximum
energy of an runaway electron $W_m$ under the action of electric
field is
$$
W_m = eE_{Dr}L,
$$
where $L$ is the characteristic length of a coronal loop.
Therefore, using Equation~(\ref{Dreicer ED}), we find
\begin{equation}
\label{lenth} L  = \frac{W}{eE_{Dr}} = W\frac{ r_{De}^2}{\alpha
e^2 \ln \Lambda} \approx 1.44 \times 10^9 \frac{W  \mbox{[eV]} T
\mbox{[K]}}{n_e \mbox{[cm]}^{-3}\ln \Lambda}.
\end{equation}
Assuming $W=100$~keV, $T = 3 \times 10^6$--$10^7$~K, $n_e =
10^8$--$10^{10}$~cm$^{-3}$, from Equation~(\ref{lenth}) we get $L
= 2.5 \times 10^9$--$6.7 \times 10^{11}$~cm. Since the
characteristic length of flare coronal loops  $L = 3 \times
10^9$~cm \citep{StepanovZaitsev19} and $n_{cr}/n_c \ll 1$ (see
Equation \ref{estimate}), the obtained estimates suggest that the
electron acceleration by sub-Dreicer electric fields seems
unlikely in solar flares \citep[see also][]{Fleishman13}. However,
we did not take into account the possible important role of
the electron acceleration by the induced electric field
for the betatron mechanism~\citep{TsapMelnikov23}. The essential
increase of the Dreicer electric field can be caused by the
ion-neutral collisions~\citep{StepanovZaitsev19} and the
interaction of accelerated electrons with turbulent
pulsations~\citep{Kaplan72}. Note that some details of the
electron acceleration by the super-Dreicer field ($E \gtrsim
E_{Dr}$) are discussed in \citep{Fleishman13}.

We used a quite rough approach for the estimates of accelerated
electrons in the ``tail'' of the Maxwellian distribution function
and did not take into consideration the Joule dissipation and
plasma heating. This should lead to a reduction of the Dreicer
field $E_{Dr}$ due to a temperature increase and, hence,  the
number of runaway electrons should  also be increased. Besides,
the Dreicer effect alone the generation of runaway electrons can
be caused by collisions between runaways and thermal electrons.
Such collisions might be infrequent, but if they do occur, there
is a high chance that after the collision both electrons will have
a velocity that is higher than the critical momentum. This
amplification of the runaway electron population is called the
avalanche mechanism \citep{Smith08}. In addition, for relativistic
runaway the friction attains a minimum value, i.e., the friction
force increases for electrons with velocities $v \approx
c$~\citep[see, e.g.,][]{GurevichZybin01}, and additional physical
effects such as radiation losses become important. These issues
need further detailed  investigations.

\begin{acknowledgements}
We would like to thank the anonymous referee for valuable remarks.
The study was supported by the Russian Foundation for Basic
Research and the Czech Science Foundation (project no.
20-52-26006, Tsap Yu.T.) and the Russian Science Foundation
(project no. 22-12-00308, Stepanov A.V. and Tsap Yu.T.).
\end{acknowledgements}

\label{lastpage}

\end{document}